# "Why I Took the Blackpill": A Thematic Analysis of the Radicalization Process in Incel Communities

Jennifer Golbeck[1], Celia Chen[1], Alex Leitch[1]

[1] University of Maryland, College Park, MD 20742
`jgolbeck@umd.edu`

**Abstract.** Incels, or "involuntary celibates", are an extreme, misogynistic hate group that exists entirely online. Members of the community have been linked to acts of offline violence, including mass shootings. Previous research has engaged with the ideologies and beliefs of incels, but none has looked specifically at the radicalization process. In this paper, we perform a thematic analysis on social media posts where incels describe their own radicalization process. We identified six major themes grouped into four chronological steps: Pre-radicalization (themes of Appearance, Social Isolation, and Psychological issues), Searching for Blame, Radicalization, and Post-Radicalization. These results align closely with existing work on radicalization among other extremist groups, bringing incel radicalization inline with a growing body of research on understanding and managing radicalization.

## 1    Introduction

The incel (involuntary celibate) community began as a community for men and women who were struggling romantically, but has since morphed into an extreme, violent misogynistic movement. Notable mass murder events include the 2014 Isla Vista killings by Elliott Rodger and the 2018 Toronto van attack by Alex Minassian. These are two of the higher profile acts of violence, but dozens of other incels have been arrested for threatening or carrying out violence.

Researchers have increasingly studied incels as an extremist group through thematic analyses of the incel experience and studies of prevalent topics including patriarchal male ideals and negative feelings about inceldom. How men radicalize from simple romantic frustration into radicalized extremists who celebrate violence and sometimes act on it remains largely unstudied through first-person accounts.

While Green et al. [1] identified a progressive "Black pill pipeline" of radicalization, their work focused on process-tracing rather than analyzing incels' own narratives of transformation. Our study addresses this gap by examining self-reports of "taking the blackpill" – the incel term for accepting their fatalistic worldview – as expressed spontaneously by community members in their natural online environments.

## 2 Background and Related Work

The Internet has transformed how extremist ideologies spread and how radicalization occurs. Research on online radicalization has established that digital environments can accelerate extremist belief adoption through several mechanisms: information provision, echo chambers, and the legitimization of extreme ideologies through reinforcement [2]. While early research on radicalization focused predominantly on Islamic extremism, recent scholarship has expanded to other ideological movements, including right-wing extremism and misogynistic communities.

Incel communities represent a form of online extremism characterized by decentralized networks that provide social belonging for isolated individuals and normalize extreme viewpoints through constant exposure and peer reinforcement [3]. Their online presence has grown substantially since 2014, following high-profile acts of violence committed by self-identified incels.

Several models explain radicalization processes. Moghaddam's [4] "staircase to terrorism" model describes progressive steps toward extremist violence, beginning with perceived injustices. Doosje et al. [5] describe a three-stage process: Sensitivity (search for significance from humiliation/status loss), Group Membership (adopting group ideology), and Action. These frameworks emphasize both push factors (grievances, identity crises) and pull factors (belonging, ideological purpose).

Recent research has specifically focused on the incel radicalization process. Green et al. [1] conducted a process-tracing analysis of the "Black pill pipeline," finding that incels utilize increasingly radical "pills" to move new members along an escalating pipeline of extremism. However, this work did not analyze incels' own narratives of their radicalization process.

Self-reported accounts of radicalization provide insights that complement observational research, often emphasizing psychological needs, social contexts, and triggering events rather than purely ideological factors. Our study addresses this gap by examining self-reports of "taking the blackpill" – the incel term for accepting their fatalistic worldview – as expressed spontaneously by community members in their natural online environments.

## 3 Method

We conducted a reflexive thematic analysis of posts from incels.is, a popular Reddit-like forum used by incels. These forums are accessible to anyone online; no account is required to read the content. The content is indexed by search engines and searchable on DuckDuckGo and Google. As previous research has noted, essentially all participants in incel forums are heterosexual men. Because users are largely anonymous, no additional reliable demographic information is available.

### 3.1 Sample

To focus on the radicalization process on incels.is, we searched for posts that contained the following phrases: "how i became an incel," "why i took the blackpill," "how i became blackpilled," "why i took the black pill," "how i took the black pill," and "became blackpilled." While this approach focused on explicit radicalization narratives rather than broader linguistic patterns, it captured users' most reflective accounts of their transformation, providing depth over breadth.

The "blackpill" is a common term used on incel forums that refers to the belief that romantic success is determined by traditionally masculine physical attractiveness, that one's genetic physical attributes cannot be meaningfully changed (e.g. by working out), and that other attributes like personality, interests, or success cannot overcome physical appearance. Not all incels are "blackpilled"; some maintain hope that they may find connection by improving themselves physically or in non-physical ways. Those incels who are "blackpilled" tend to regard themselves as the most extreme or dedicated in the community. Because our research question focuses on radicalization, this is a useful marker.

We searched for these terms both on Google using the site:incels.is filter and in the Radicalization and Deradicalization in Online Communities dataset of Incel forums [6]. It is common for users of these forums to share stories of how they became incels or how they became blackpilled, so this sampling method generated a large number of threads. We identified 60 forum posts with the above search terms and collected the original posts and replies, since the replies often included personal stories of radicalization. This resulted in 73,614 words of text. In performing the thematic analysis, we only considered posts and replies that discussed the personal radicalization process; we did not code comments about other issues.

### 3.2   Data Analysis

We used Braun and Clarke's [7] six-phase thematic analysis process oriented around: "How do posters describe their radicalization?" After data familiarization, we generated initial codes. Three researchers (one professor and two graduate students) coded all posts, with each analyzed for radicalization process content. Posts ranged from one sentence to multiple paragraphs, with multiple codes possible per post.

All researchers had previous experience coding incel data [8] and were familiar with an incel lexicon built on this dataset [9]. Using Braun and Clarke's reflexive approach, we reached consensus on code meanings through discussion rather than formal inter-rater reliability. After initial coding, researchers synthesized codes into themes, with a natural chronological order emerging as the organizing structure.

## 4   Results

Six themes emerged from this analysis. The thematic map is shown in Figure 1. The themes are presented in chronological order, reflecting the radicalization process described in the posts we reviewed. The first chronological category, which contains three themes, represents the pre-radicalization phase where these posters discuss the difficult conditions that existed in their lives before they began the radicalization process. The second chronological category, with Theme 4, represents searching for something or someone to blame for the difficulties of their lives in the pre-radicalization phase. The third chronological category, with Theme 5, describes the process of becoming radicalized, and the fourth, with Theme 6, describes feelings in the post-radicalization state.

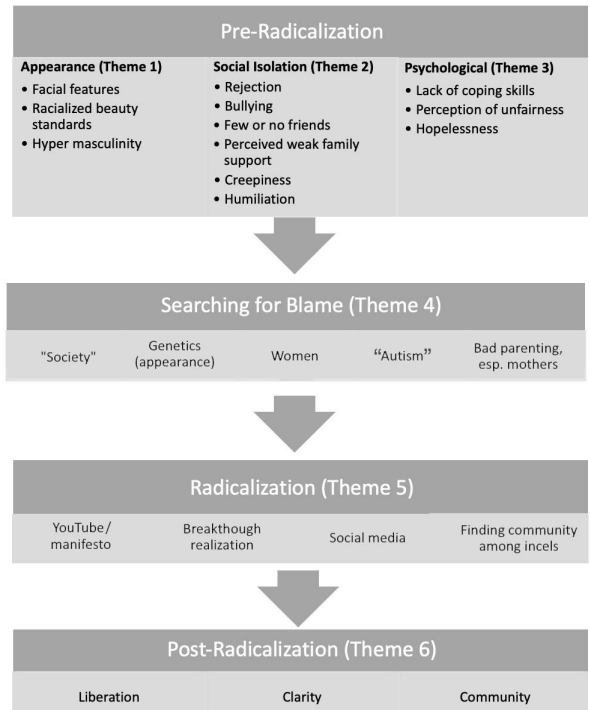

*Figure 1. Thematic map with the six themes organized in a chronological framework*

## 4.1   Pre-Radicalization – Themes 1-3

Before discovering incel communities or becoming radicalized, posters describe difficult conditions in their lives. There are three themes: Appearance (Theme 1), Social Isolation (Theme 2), and Psychological (Theme 3). The first two are explicitly named by posters as conditions that were difficult in their lives. The third psychological theme is more implicit in discussions of situations they struggled with. Together, these themes describe the groundwork that leads to the radicalization path.

**Theme 1: Appearance.** The core of the "blackpill" is that only men meeting specific, high standards of physical attractiveness can be successful with women. Posters describe feeling insecure and rejected because of their appearance before discovering incel content. These are feelings they describe having before discovering blackpill-related content—they believed their appearance was part of their problem and that inspired their search for a reason why.

One poster wrote: "I mean I always knew I was fucked up, ever since age 10-11 I knew my life will be fucked. But I wasn't 'blackpilled' back then, I didn't know about the blackpill space or lingo... I knew it was to do with me being ugly and me being an autist." Another wrote, "Growing up ugly, I understood the significance of looks at age 11."

Racialized views of attractiveness also appear frequently. In these communities, traditionally Aryan features (tall, sharp jawline, white or paler skin, blonde hair, blue eyes) are considered most attractive and often necessary conditions to find a romantic connection. Posters describe recognizing their race/non-conformity with this standard as a problem early on.

**Theme 2: Social Isolation.** Posters describe social isolation from a young age through bullying or social rejection. One wrote, "I was bullied in over 5 schools throughout my life so I knew I was a cursed soul." Another said, "When I was 11, I realised that foids [women] would only reject me and they would laugh at me."

Posters often connect these problems to their appearance: "constantly being emotionally tortured by the people around me and just not fitting in normie groups... and also not having the looks and height that I desire." Social isolation is well-established as an important factor in radicalization [10].

**Theme 3: Psychological.** In their discussions, posters reveal a number of psychological factors that prime them for radicalization. They display a lack of coping skills, and as research predicts, this leads to expressions of hopelessness, high reactivity, and aggression to attacks.

When explaining why he became radicalized, one poster described an incident that he found humiliating: "[I] always knew. final nail in the coffin was gym class 7th grade where we were forced to learn square dance or some shit. teacher told hot bitch to hold my hand and she did that thing with her sleeve as if she was touching trash. was forced to stare her down for 40 minutes." In this scenario, a girl covered her hand with her sleeve and the poster found this so difficult to cope with that he reacted with aggression, staring her down, and continues lashing out in anger at the time of posting, using slurs to describe the girl and claiming that this incident from years earlier was the final incident that radicalized him.

## 4.2 Searching for Blame (Theme 4)

After the pre-radicalization phase creates conditions where incels feel rejected and hopeless, the next step involves forsaking responsibility and seeking external blame targets including society, women, genetics, autism, and bad parenting. One poster wrote, "The Blackpill is merely a recognition of the reality that for guys like us there is nothing that can be done to change our situation. It's the fault of our parents and society writ large."

Autism is a persistent target of blame. One poster wrote, "I knew it was to do with me being ugly and me being an autist." This blame-seeking creates cognitive openings that prime individuals for radicalization.

## 4.3 Radicalization (Theme 5)

Theme 5 captures the actual radicalization process. Up to this point, incels describe isolation, social rejection, difficulty coping, and struggles based on their appearance.

They look for forces to blame, then explain the moments in which they were radicalized or "took the blackpill".

For many, their radicalization was catalyzed on YouTube through channels like Wheat Waffles, FACEandLMS, HeedandSucceed, and Incel TV. One poster wrote, "One day i found a comment... saying 'FaceandLMS was right'... I just watched the WAW Videos and i did not know whether i should cry or laugh... The WAW ['What Attracts Women'] videos where a like a blackpill-nuclear-bomb dropped on me."

Others describe breakthrough realizations: "One time a foid [female] coworker... gave it to me with a nasty face. Then my coworker arrived and she was blushing... That was my eureka moment. Suddenly it made sense why women... would act mean towards me."

### 4.4 Post-Radicalization (Theme 6)

After radicalizing by "taking the blackpill," accepting that they will never achieve romantic success because they do not meet the unfair standards and expectations of society, posters describe their feelings in the post-radicalization state. These generally describe feelings of liberation, clarity, and community.

One poster explains how the blackpill allows him to live without taking responsibility for his situation, writing, "Being blackpilled means you don't blame yourself for any of it which is liberating." Another writes, "The Blackpill is merely a recognition of the reality that for guys like us there is nothing that can be done to change our situation. It's the fault of our parents and society writ large, that may sound a bit pretentious to the average normie but it's the truth."

The blackpill is essentially a radicalization into hopelessness. Previous research has linked hopelessness with risk of suicide. Suicide is an extremely common topic of discussion on incel forums. Hopelessness has also been linked to violence in many contexts [11]. Though most incels do not carry out acts of violence, they are responsible for a number of high-profile incidents, including mass murders. Incel communities celebrate these acts – and other mass shootings carried out by non-incels.

## 5 Discussion

The findings in this study echo more general findings on the radicalization process. In studying radicalization among terrorists, Doosje et al. [5] describe a three-stage process: Sensitivity, Group Membership, and Action. Sensitivity describes a search for significance, which the authors describe as arising from feelings of a loss of status or strong sense of humiliation. These are captured clearly in Themes 1-3. This creates a cognitive opening, which we see in Theme 4, the pre-radicalization step. The Group Membership phase involves joining a group where the individual adopts the group's ideology and shows loyalty to the group. We find this in Theme 5, the radicalization process where incels "take the blackpill", and in Theme 6 where they find relief in being part of the blackpilled ideology.

Our results also echo research which found that the internet provides a space for effective and inexpensive communication of radical ideas, paired with anonymity, that lets those seeking information to be recruited and become radicalized. Most of our posters, when discussing where they discovered the "blackpill" mention social media broadly and YouTube specifically as a source of information.

This alignment with theories and prior results suggests that incels follow a fairly typical psychological radicalization process and that work on interventions and deradicalization can be carried over into incel community analysis. The role of social media, particularly YouTube's recommendation algorithms, in facilitating radicalization parallels findings in other extremist contexts where users progress from mainstream to increasingly extreme content.

### 5.1 Limitations and Future Work

Our dataset draws from a single forum, limiting generalizability. Future research should examine multi-platform radicalization pathways and employ network analysis approaches to understand interaction dynamics and content diffusion patterns within these communities. Additionally, quantitative analysis of blame targets and emotional vs. cognitive appeals in radicalization narratives could inform targeted intervention approaches.

While this paper deals with social media discourse, future research should incorporate network analysis approaches to understand structural dynamics, interaction patterns, and user trajectories—elements central to social network analysis and mining. A graph-based analysis of interaction networks or diffusion patterns of radical narratives would provide valuable insights into how radicalization spreads through these communities.

Additionally, our search methodology, while capturing explicit radicalization narratives, may have missed users who describe their transformation using alternative phrases. Future work could employ more systematic data collection methods and complement qualitative analysis with quantitative approaches to understand the main targets when incels blame others and how emotion versus cognition differently influence the radicalization process.

## 6 Conclusion

Using a set of posts where incels self-described their radicalization process, we conducted a thematic analysis and identified six major themes with four chronological steps: Pre-radicalization (themes of Appearance, Social Isolation, and Psychological issues), Searching for Blame, Radicalization, and Post-Radicalization. In this data, men describe struggling to form relationships early in childhood, often blaming their appearance for the bullying and rejection they experienced. They also reveal extreme emotional reactions to fairly typical social awkwardness and a difficulty coping. This is followed by an abdication of responsibility and search for someone or something to blame. Radicalization occurs when they "discover" the blackpill, an ideology that prescribes that only hyper-masculine men who meet a certain attractiveness standard can find romantic relationships with women. They usually discover this ideology through social media, particularly on YouTube. After "taking the blackpill" and radicalizing, they describe feelings of peace, clarity, and community membership within blackpill forums, though these are often accompanied by angry posts and violent fantasies.

Our results align with many existing studies on radicalization in other communities. This suggests that incels follow a somewhat typical pattern of radicalization with the same social and psychological factors at play. This lays the foundation to allow future work on interventions and deradicalization to be translated to inceldom. Similarly, ongoing work on deradicalizing incels that is arising from online forums like /r/IncelExit that support men leaving the community, may offer insights into supporting deradicalization of extremists in other contexts.

The research presented here builds upon existing work on incel communities by applying thematic analysis to understand the radicalization process as described by incels themselves. By analyzing personal narratives of how individuals became radicalized, this study contributes to the growing body of knowledge on online extremism, with implications for intervention strategies and prevention efforts. Future research incorporating network analysis and multi-platform approaches will further enhance our understanding of how radicalization spreads through online communities and identify optimal intervention points.